\documentclass[10pt,twocolumn]{IEEEtran}
\pdfoutput=1
\usepackage{cite}
\ifCLASSINFOpdf
   \usepackage[pdftex]{graphicx}
   \DeclareGraphicsExtensions{.pdf,.jpeg,.png}
\else
   \usepackage[dvips]{graphicx}
   \DeclareGraphicsExtensions{.eps}
\fi
\usepackage{amssymb}
\usepackage{upgreek}
\usepackage{latexsym}
\usepackage{multirow}
\usepackage{epstopdf}
\usepackage{algpseudocode}
\usepackage{xcolor}

\usepackage[cmex10]{amsmath}
\usepackage{amssymb}
\usepackage{array}
\usepackage{mdwmath}
\usepackage{mdwtab}
\usepackage{pslatex}
\usepackage{color}
\usepackage{textcomp}
\usepackage{placeins}
\usepackage{eqparbox}
\usepackage{url}
\usepackage{stfloats}
\usepackage{multicol}
\usepackage{subfigure}
\usepackage{amsthm}
\usepackage{comment}
\usepackage{balance}
\usepackage{gensymb}

\begin{document}

\title{\vspace{-5mm}Breaking the Limits in Urban Video Monitoring:\\Massive Crowd Sourced Surveillance over Vehicles}

\author{Vitaly Petrov, Sergey Andreev, Mario Gerla, and Yevgeni Koucheryavy\vspace{-8mm}
\thanks{V. Petrov, S. Andreev, and Y. Koucheryavy are with Tampere University of Technology, Finland.}
\thanks{Y. Koucheryavy is also with National Research Institute Higher School of Economics, Moscow, Russia.}
\thanks{M. Gerla is with University of California, Los Angeles (UCLA), USA.}
}

\maketitle

\begin{abstract}
Contemporary urban environments are in prompt need of means for intelligent decision-making, where a crucial role belongs to smart video surveillance systems. While existing deployments of stationary monitoring cameras already deliver notable societal benefits, the proposed concept of massive video surveillance over connected vehicles that we contribute in this paper may further augment these important capabilities. We therefore introduce the envisioned system concept, discuss its implementation, outline the high-level architecture, and identify major data flows, while also offering insights into the corresponding design and deployment aspects. Our conducted case study confirms the potential of the described crowd sourced vehicular system to effectively complement and eventually surpass even the best of today's static video surveillance setups. We expect that our proposal will become of value and integrate seamlessly into the future Internet-of-Things landscape, thus enabling a plethora of advanced urban applications.
\end{abstract}

\IEEEpeerreviewmaketitle

\vspace{-5mm}
\section{Introduction}
\label{sec:intro}

The share of video surveillance traffic is to become around $3.9\%$ of the global Internet traffic in 2020, which equals to astounding $1.3$ exabytes per month\footnote{Cisco, ``The Zettabyte Era: Trends and analysis,'' \emph{White Paper}, 2016. \url{https://www.cisco.com/c/en/us/solutions/collateral/service-provider/visual-networking-index-vni/vni-hyperconnectivity-wp.html}, Accessed: June 2018.}. This includes data from numerous applications, such as mass event monitoring and facility protection, public safety and crime deterrence, outdoor perimeter security and road traffic control, and many more~\cite{ref_video}. The importance of large-scale video surveillance and the prospective benefits that it brings to different economy sectors transform video surveillance systems and services into a multi-billion market, which grows annually around $15\%$~\cite{IHS_Markit}.

However, even in the most developed cities, such as London, New York, and Singapore, the penetration of stationary surveillance cameras remains limited. The reasons are rooted in economic and social factors, since video surveillance cameras may create a sense of personal discomfort~\cite{discomfort_book}. At the same time, such stationary systems with only partial coverage -- being a typical case in today's city-scale deployments -- have inherent limitations when monitoring `mobile' events. The currently observed moving vehicle may e.g., turn around a corner and disappear from sight. Hence, static systems can hardly capture all of the events of interest as long as they rely solely on stationary cameras. 


Meanwhile, contemporary private cars equipped with capable \textit{vehicular} cameras not only have high penetration in urban environments but also demonstrate better diversity as compared to stationary surveillance systems. The massive utilization of vehicular cameras can potentially \emph{complement existing stationary video surveillance} in the public safety systems of tomorrow. The corresponding video captured by such moving cameras constitutes a new source of massive data, calls for new analysis techniques, and aims to rapidly improve the capabilities of next-generation surveillance systems. However, city-scale video collection from moving vehicular cameras imposes numerous research challenges and system design questions that need to be addressed carefully.

The crucial role of video capture, exchange, and analysis in vehicular contexts has been noted in various sources, including the recent 5G-PPP documents\footnote{5G-PPP, ``5G Automotive Vision,'' \emph{White Paper}, October 2015. \url{https://5g-ppp.eu/wp-content/uploads/2014/02/5G-PPP-White-Paper-on-Automotive-Vertical-Sectors.pdf} Accessed: June 2018.}. At the same time, most proposals concerning connected vehicles for video exchange are primarily focused on commuting-centric applications~\cite{heath_vehicular_sensing}, as well as traffic safety~\cite{related_work2} and collision avoidance systems~\cite{related_work3}. In other words, the automotive sector is often assumed to collect, transfer, and process video data for its internal use but is not expected to produce any notable benefits to other verticals. In rare cases where vehicular assistance was considered, it is mainly limited to in-car video surveillance~\cite{ref_camera} and video delivery for in-vehicle entertainment systems\footnote{Qualcomm, ``Leading the world to 5G: Cellular Vehicle-to-Everything (C-V2X) technologies'', \emph{White Paper}, June 2016. \url{https://www.qualcomm.com/media/documents/files/cellular-vehicle-to-everything-c-v2x-technologies.pdf} Accessed: June 2018.}.

Hence, the benefits of enabling massive outdoor video surveillance over vehicles have not been studied sufficiently, while attractive design options and associated research challenges for the underlying communications infrastructure have not been reviewed systematically. In this work, we target to fill-in this gap by proposing a novel concept of \textit{crowd sourced video surveillance over vehicles}. We thus outline the overall system architecture and implementation choices, discuss the technical challenges instrumental to its successful deployment, and conduct a case study for the envisioned system by assessing its capability to capture different categories of characteristic events. The described concept introduces a novel use case for high-rate vehicular communications, which is featured by rich cross-sectoral cooperation and direct monetization of the video data.



\vspace{-3mm}
\section{Our Proposed Concept}
\label{sec:concept}

\subsection{Vehicles as Mobile Cameras}

\subsubsection{The underlying idea}

In this paper, we propose a non-conventional approach to obtain massive amounts of video data and combine/process them to produce valuable knowledge, which is based on employing connected vehicles. Our introduced crowd sourced \emph{Video Surveillance over Vehicles (VSV)} system can be utilized to efficiently collect, combine, and process a large number of multimedia streams coming from various vehicular cameras\footnote{We consider both built-in cameras and dashboard cameras (dashcams). For simplicity, further we primarily refer to built-in solutions, whereas our approach is also applicable for dashcams with minimal modifications.}. Our approach to VSV system design is illustrated in Fig.~\ref{fig:idea}, where almost any location on or next to the road may be observed by a number of passing vehicles. From the practical perspective, a new road traffic ecosystem stakeholder -- the VSV service provider -- is introduced, thus complementing the already existing fleet of connected vehicles, the network operators, and the service consumers.

\subsubsection{Technology enablers are in place}

The rationale behind the proposed vision stems from the fact that modern vehicles already are or will soon be equipped with: (i) high-resolution sensing devices (cameras, radars, LiDARs, etc.) as well as intelligent recognition techniques to detect pedestrians, road signs, etc.; (ii) positioning capabilities via satellite navigation systems, and (iii) high-rate connectivity modules. Hence, all of the technology enablers are already in place or will be available soon. Moreover, the vehicles of tomorrow are envisioned to be increasingly intelligent entities. When interconnected into a \emph{vehicular cloud network}~\cite{mario_vehicular_cloud}, they can offer benefits to other sectors, thus becoming an integral part of the 5G-grade Internet of Vehicles. Our proposed concept makes a step ahead w.r.t. the state-of-the-art considerations and conceptualizes an example crowd sourced video surveillance service.

\vspace{-3mm}
\subsection{Key System Design Considerations} \label{subsec:stationary_vs_mobile}

There are many existing techniques for video surveillance over stationary cameras (e.g., object recognition). These can be applied to the proposed VSV system with minimal changes already today. However, the mobile and more dynamic nature of crowd sourced video traffic implies several new challenges in the respective system design. 

\subsubsection{What to collect?}

Altogether, there are four data streams to be captured by the VSV-capable car. The first two are the video and time streams coming from the camera and used to perform the actual monitoring. In addition, the stream with the vehicle's location and the ``compass'' stream are needed to determine the current moving direction. As the ``location'' stream is retrieved from e.g., satellite positioning systems, it has to be fed with the data coming from the navigation system or a built-in compass. The combination of the latter two streams allows to accurately bind the coverage of a moving camera to a certain geographical area. The total volume of the collected data is thus similar to that from static cameras, since the video stream size still dominates in terms of bitrate.

\begin{figure}[!t]
  \centering
  \includegraphics[width=1.0\columnwidth]{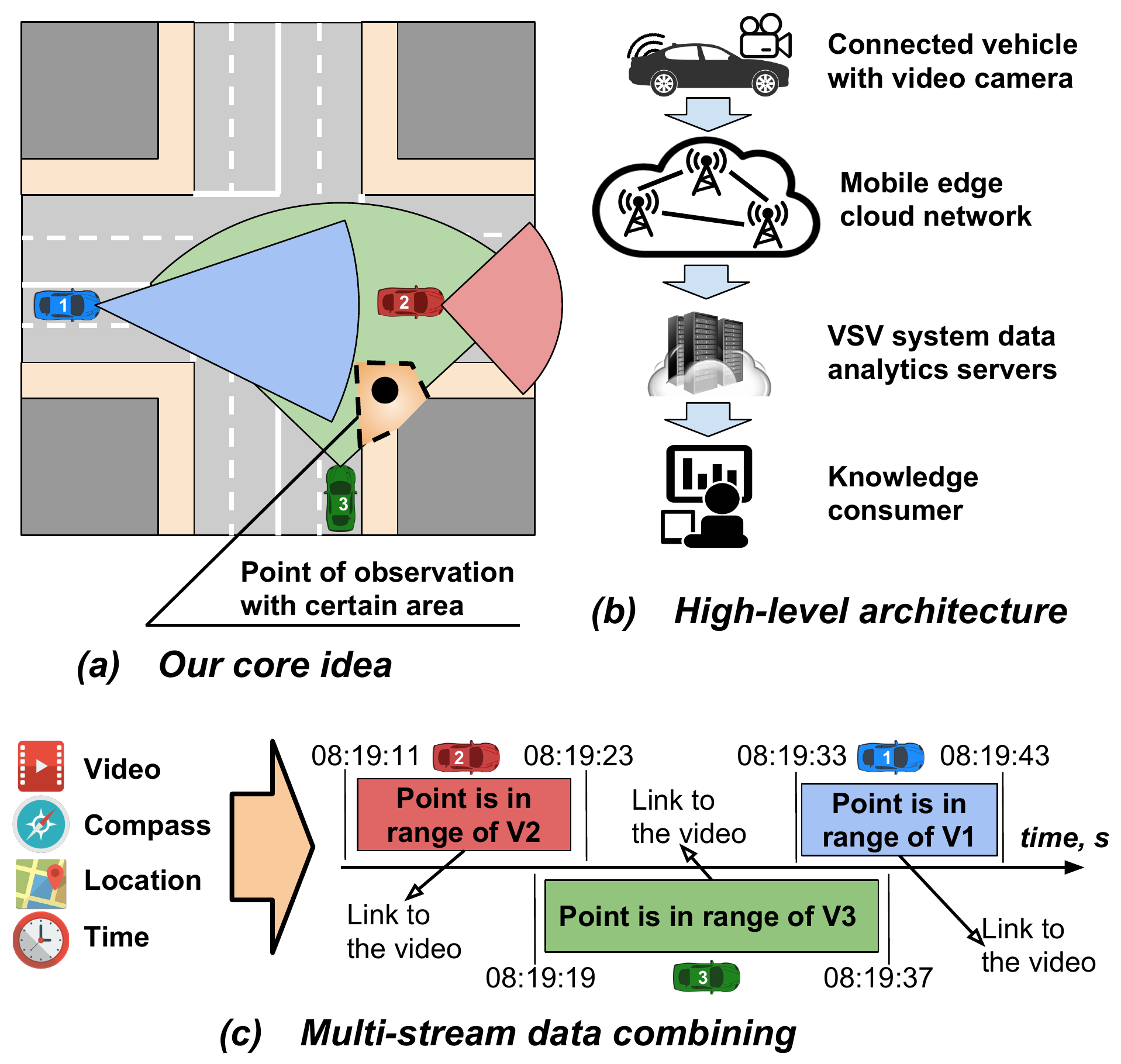}
  \vspace{-5mm}
  \caption{Our approach to crowd sourced video surveillance over vehicles.}
  \label{fig:idea}
  \vspace{-5mm}
\end{figure}

\subsubsection{How to deliver?}

Data exchange between the connected vehicles and the network infrastructure is a challenging research issue. As some of the technology insights into the VSV data delivery are outlined in the following section, the key consideration is that a closer cooperation between the network operator and the VSV system provider is required. The former has to be made aware of the deployed VSV system and its parameters. In its turn, the VSV service provider should be informed about the properties of the network deployment as well as receive assistance from the network operator (e.g., instructions to postpone current transmission until the vehicle leaves a congested area).

\subsubsection{How to combine?}

While stationary cameras may monitor a certain area for longer durations of time, moving cameras in the VSV system pass rapidly from one place to another. Consequently, the video stream associated with the area of interest will comprise several (overlapping) fragments received from different vehicles, as illustrated in Fig.~\ref{fig:idea}(c). The following four-stage procedure is proposed to convert the initial set of `vehicle-associated' video streams into the conventional and more convenient set of `area-associated' streams: 

$\rhd$ \underline{\emph{Stage 0.}} At the configuration stage, the entire area under surveillance is fragmented into smaller zones, each represented by a certain \emph{point of observation} in the proposed VSV system.

$\rhd$ \underline{\emph{Stage 1.}} The localization and compass streams are used to associate each moment of time with the points of observation that are currently within the camera's range. Here, the video stream can also be utilized to cross-check this association by recognizing known objects in the video.

$\rhd$ \underline{\emph{Stage 2.}} The association data is combined with the time stream in order to identify the start and the end moments of time, when the video stream at hand is relevant to the selected point of observation.

$\rhd$ \underline{\emph{Stage 3.}} The calculated time intervals are then used to virtually fragment the video file into the truncated pieces, each related to its own point of observation (e.g., point in Fig.~\ref{fig:idea}(c) is observed by Vehicle 3 from 08:19:19 to 08:19:37).

$\rhd$ \underline{\emph{Stage 4.}} The truncated virtual fragments are stored next to the previously received fragments related to the same point of observation. The time intervals are employed to properly combine the fragments into a single resulting stream.

\begin{figure}[!t]
  \centering
  \includegraphics[width=1.0\columnwidth]{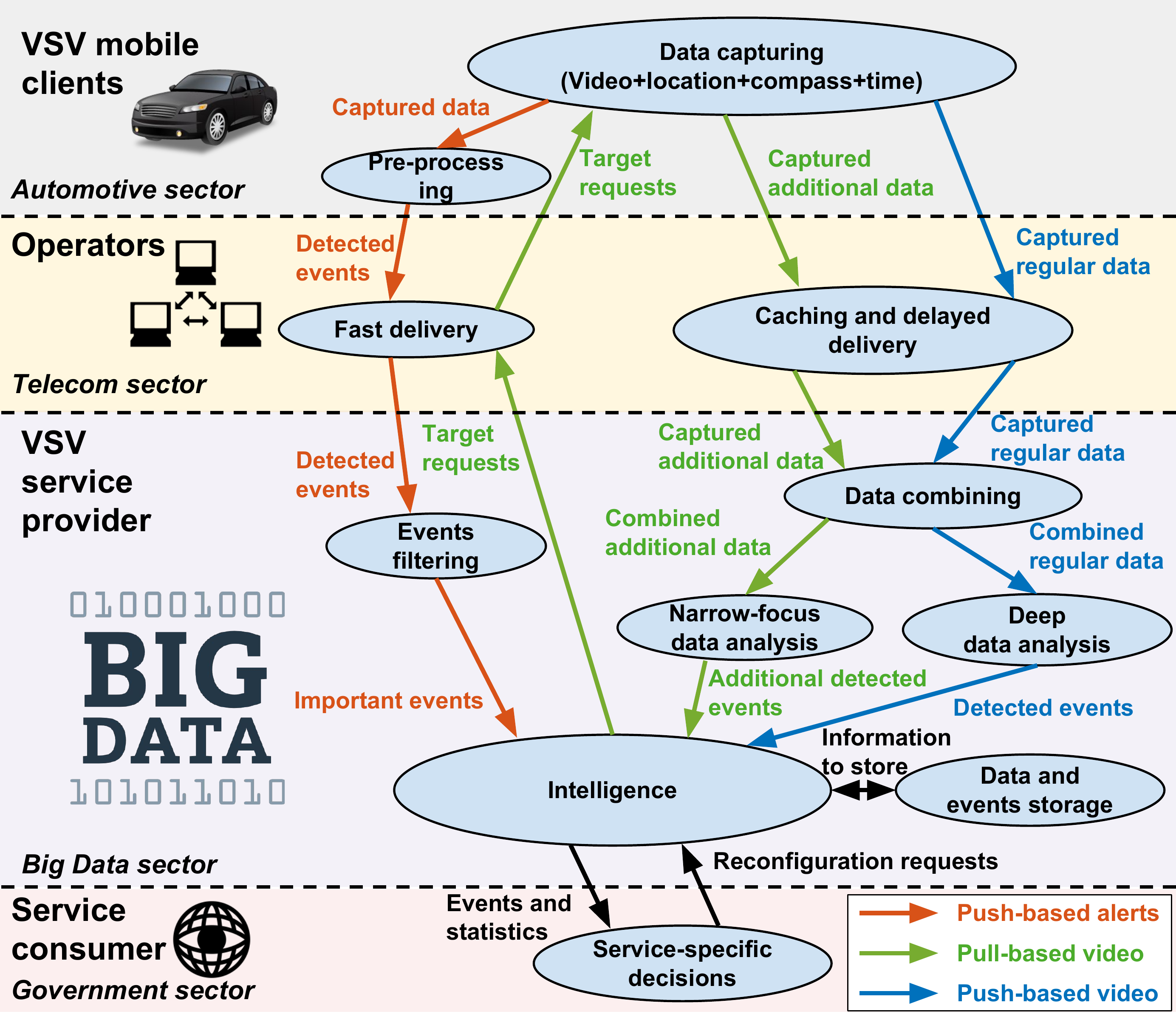}
  \vspace{-5mm}
  \caption{High-level architecture and key data flows in our VSV system.}
  \label{fig:architecture}
  \vspace{-5mm}
\end{figure}

\vspace{-3mm}
\subsection{High-Level Architecture and Data Flows}

\subsubsection{Participating stakeholders and their roles}

Construction of a practical VSV system calls for close multi-disciplinary collaboration between the various stakeholders. Particularly, four major actors are envisioned, who represent automotive, telecommunication, Big Data analytics, and government clusters. The proposed system architecture and the roles of these sectors are illustrated in Fig.~\ref{fig:architecture}.

The automotive sector is responsible for capturing, pre-processing, and storing the data locally, as well as for facilitating information delivery to the access network. The role of the network operator is to provide efficient mechanisms for collecting the data streams from the connected vehicles as well as for having them delivered to the VSV service provider. The latter offers the key functionality in our envisaged system by combining the heterogeneous data streams received from different network operators, storing them, and ultimately processing such accumulated data to detect certain events as well as produce useful knowledge. The results of the said analytics are later converted into the form convenient for the consumers, thus helping them make informed decisions. The consumers may also interact with the VSV system directly by issuing certain configuration requests e.g., asking to run advanced algorithms for the information coming from a certain area.

\subsubsection{Diversity of data flows}

Depending on the target application and the area of interest, the VSV system may utilize various data flows. In Fig.~\ref{fig:architecture}, we introduce a set of three flows that is sufficient for the baseline system design.

\begin{itemize}
\item \textbf{Push-based video.} This is a generic flow of data continuously collected by the vehicle. It is a logical equivalent of the conventional surveillance system based on the stationary cameras. The set of relevant applications ranges from real-time area monitoring to backtracking of the occurred crime.

\item \textbf{Push-based alerts.} The purpose of this flow is to process as much data as feasible locally in the vehicle. This data flow can be utilized to search for certain predefined events/objects e.g., locate missing people/pets, criminals (via face/silhouette recognition), or stolen vehicles (via number plate recognition).

\item \textbf{Pull-based video.} This data flow pertains to the use cases where a certain area/object/event attracted attention of the VSV system (or its customer) and thus further source data are required. Here, the VSV system may send targeted requests to the nearby vehicles demanding them to switch on the video capturing device or even drive into the area of interest if the vehicle is currently nearby. Appropriate car owner involvement and incentivization mechanisms need to be provided for the latter option~\cite{mario_incentives}.
\end{itemize}

\vspace{-1mm}
\section{Detection of Mobile Events -- A Case Study} \label{sec:numerical}

The deployment of the described VSV system may require certain effort by various stakeholders. Hence, the potential gains of its implementation should be made clear. To this aim, we compare the performance of the proposed VSV system with that of the reference video surveillance setup utilizing stationary cameras. For a fair comparison between (mostly) wired static surveillance deployments and the VSV system relying on wireless access, we provision the connectivity of vehicle-mounted cameras, such that any captured video fragment is delivered reliably to the VSV service provider. The goal of our study is to understand the extent of benefit that VSV offers from the service perspective before studying any possible bottlenecks in the access network handling the VSV video traffic. Our below case study confirms that the VSV system can efficiently complement the already deployed stationary solutions in capturing various types of mobile events.

\vspace{-3mm}
\subsection{Our Setup and Framework}

\subsubsection{Categories of events}

The features of the observed event, such as its total duration and evolution in space, may notably affect the performance results. To offer a balanced judgment, this study considers both short- and long-term events that `move' at different speeds, as illustrated in Fig.~\ref{fig:case_study}(a). Altogether, four event categories are addressed:

$\rhd$ \emph{`Explosion'.} This event category reflects numerous cases where an explosive detonates in the middle of a city, caused by an accident (e.g., gas detonation) or a terrorist attack. The considered event category is characterized by short duration and no mobility. Car accidents, frequently appearing in urban traffic (e.g., on a crossroad), have similar characteristics.

$\rhd$ \emph{`Picket'.} This event category reflects a small-scale demonstration featuring several tens of people. Each such event lasts from tens of minutes to several hours and remains static or moves with the walking speed.

$\rhd$ \emph{`Robbery'.} This event category corresponds to a street robbery incident where a criminal e.g., snatches a bag and runs away. In contrast to the previous category, this event is typically characterized by higher mobility (a running human) and shorter duration.

$\rhd$ \emph{`Vehicle'.} This event category is reflective of police chasing a vehicle with suspects. The action may last for quite a while in time as well as move from one side of the city to another at the vehicular speed. As a result, this event is generally one of the most challenging for the video surveillance systems to capture.

\subsubsection{Metrics of interest}

The efficiency of the considered video surveillance system in monitoring certain categories of events can be well characterized by the following set of metrics:

\begin{enumerate}
\item \textbf{Probability of event detection}. This parameter determines the chances to observe even a short part of the target event. In this study, we define \textit{detection} as a case when at least one square meter of the event area was monitored for at least one second of time, since smaller shares across either time or space might not result in successful event recognition.
\item \textbf{Probability of event monitoring}. With this parameter, we evaluate the chances that a certain event was monitored continuously by at least one camera for its entire duration. This is much more difficult to achieve, especially for longer events moving at higher speeds.
\item \textbf{Fragmentation of video stream}. We also assess the distribution of the number of different video fragments that need to be combined by the VSV system in order to produce the stream associated with a certain area.
\end{enumerate}

While the first two metrics aim to characterize the overall system usability in terms of event detection and monitoring, the third one provides insights into how complex the combining process will be.

\subsubsection{Deployment parameters}

\begin{figure}[!t]
  \centering
    \vspace{-7mm}
  \includegraphics[width=0.85\columnwidth]{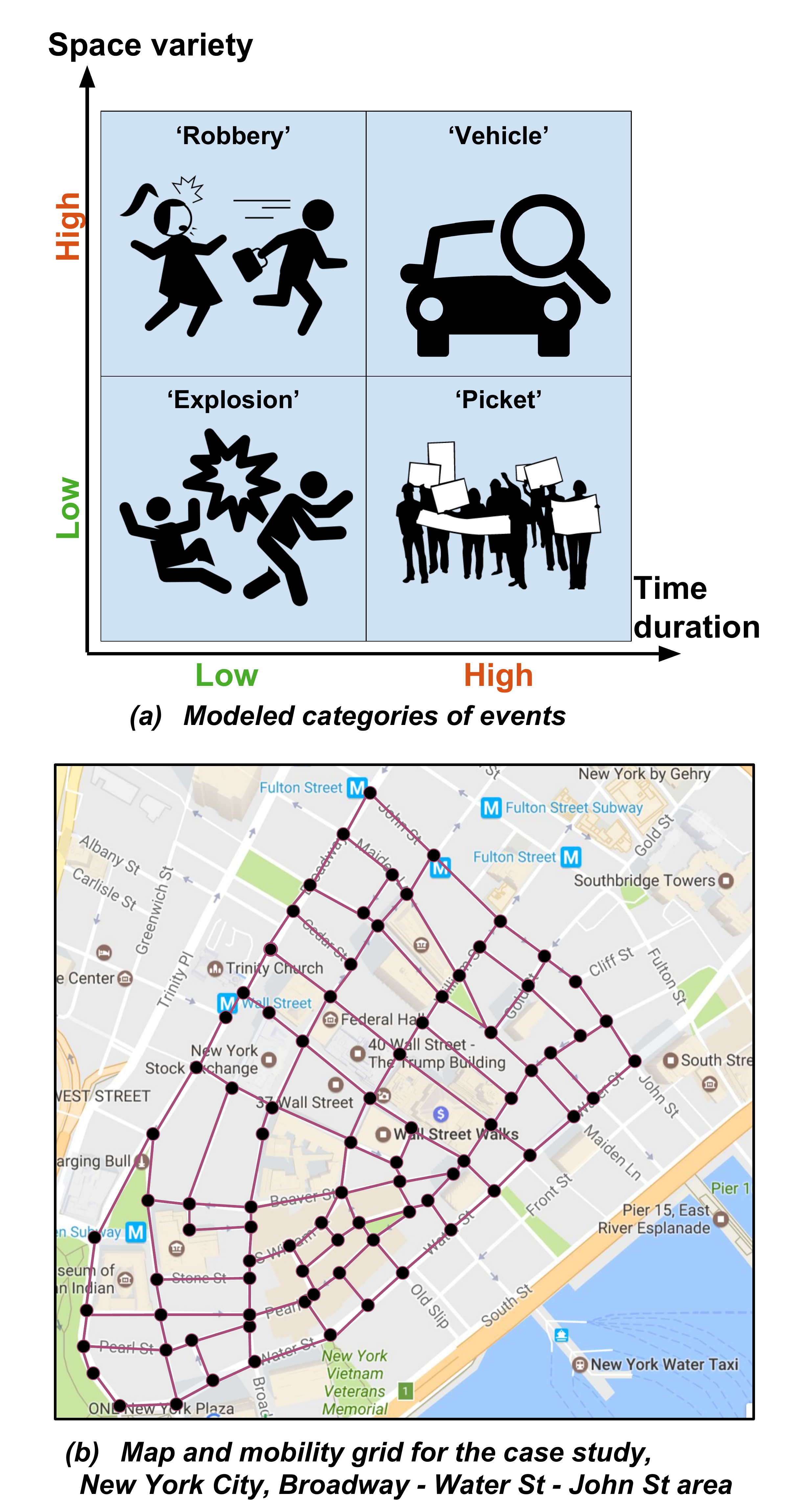}
  \vspace{-3mm}
  \caption{Event categories and setup for our case study.}
  \vspace{-7mm}
  \label{fig:case_study}
\end{figure}

In what follows, we consider a typical urban deployment. More specifically, a part of the Lower Manhattan is chosen, as illustrated in Fig.~\ref{fig:case_study}. For simplicity, we do not model pedestrians in our scenarios except for those directly involved into the observed events. For the same reason, two-way traffic is assumed on all of the streets. Altogether, there are around $2,000$ vehicles per sq. km in the considered setup\footnote{Estimated using Google Street View as approximately one vehicle per $20$\,m of the road.}. The \emph{penetration of the VSV system}, which is defined as a share of vehicles contributing to our service, is a parameter that varies from $10\%$ (low penetration) to $100\%$ (full penetration). To compare the proposed system with other existing solutions, stationary cameras are considered as well. All of the cameras are deployed uniformly along the streets with a certain density.

\begin{table}[b!]
\vspace{-2mm}
{\small
\renewcommand{\arraystretch}{1}
\caption{Case study simulation parameters.}
\vspace{-3mm}
\label{tab:parameters}
\begin{center}
\begin{tabular}{p{0.44\columnwidth}p{0.46\columnwidth}}
\hline
\textbf{Parameter}                & \textbf{Value}   \\
\hline\hline
\multicolumn{2}{c}{\emph{Video Surveillance over Vehicles (VSV) system}}\\
\hline
Density of vehicles &1,967 per sq. km\\
\hline
Vehicle camera range / angle & 30\,m / 120$\degree$\\
\hline
Vehicle mobility pattern & Manhattan grid, 15\,m/s speed\\
\hline
\multicolumn{2}{c}{\emph{Reference stationary video surveillance system}}\\
\hline
Stationary camera range & 50\,m / 120$\degree$\\
\hline
\multirow{3}{*}{Density of cameras} & 300 per sq. km (London)\\
& 166 per sq. km (New York City)\\
& 123 per sq. km (Paris)\\
\hline
\multicolumn{2}{c}{\emph{Event-specific parameters}}\\
\hline
`Explosion' time / area / speed & 2\,s / 2\,sq. m / stationary\\
\hline
`Picket' time / area / speed & 10\,min / 100\,sq. m / 1\,m/s\\
\hline
`Robbery' time / area / speed & 10\,s / 1\,sq. m / 5\,m/s\\
\hline
`Vehicle' time / area / speed & 30\,min / 8\,sq. m / 15\,m/s\\
\hline
Event mobility pattern & Manhattan grid, given speed\\
\hline
\end{tabular} 
\end{center}
}
\end{table}

The density of stationary cameras is another parameter that varies from $123$ per sq. km (reflecting a video surveillance system in Paris\footnote{Defence One, ``Here's why security cameras were no help in capturing Paris terrorists'', January 2015. \url{http://www.defenseone.com/threats/2015/01/heres-why-security-cameras-were-no-help-capturing-paris-terrorists/102438/} Accessed: June 2018.}) through $166$ per sq. km (the case of New York City\footnote{Reuters, ``NYPD expands surveillance net to fight crime as well as terrorism,'' June 2013. \url{http://www.reuters.com/article/usa-ny-surveillance-idUSL2N0EV0D220130621} Accessed: June 2018.}), and up to $300$ per sq. km as in London, which is currently one of the most monitored cities in the world\footnote{UrbanEye ``CCTV in London,'' \emph{Working paper}, June 2002. \url{http://www.urbaneye.net/results/ue_wp6.pdf} Accessed: June 2018.}.

\begin{figure*}[!t]
  \centering
  \vspace{-8mm}
  \includegraphics[width=1.0\textwidth]{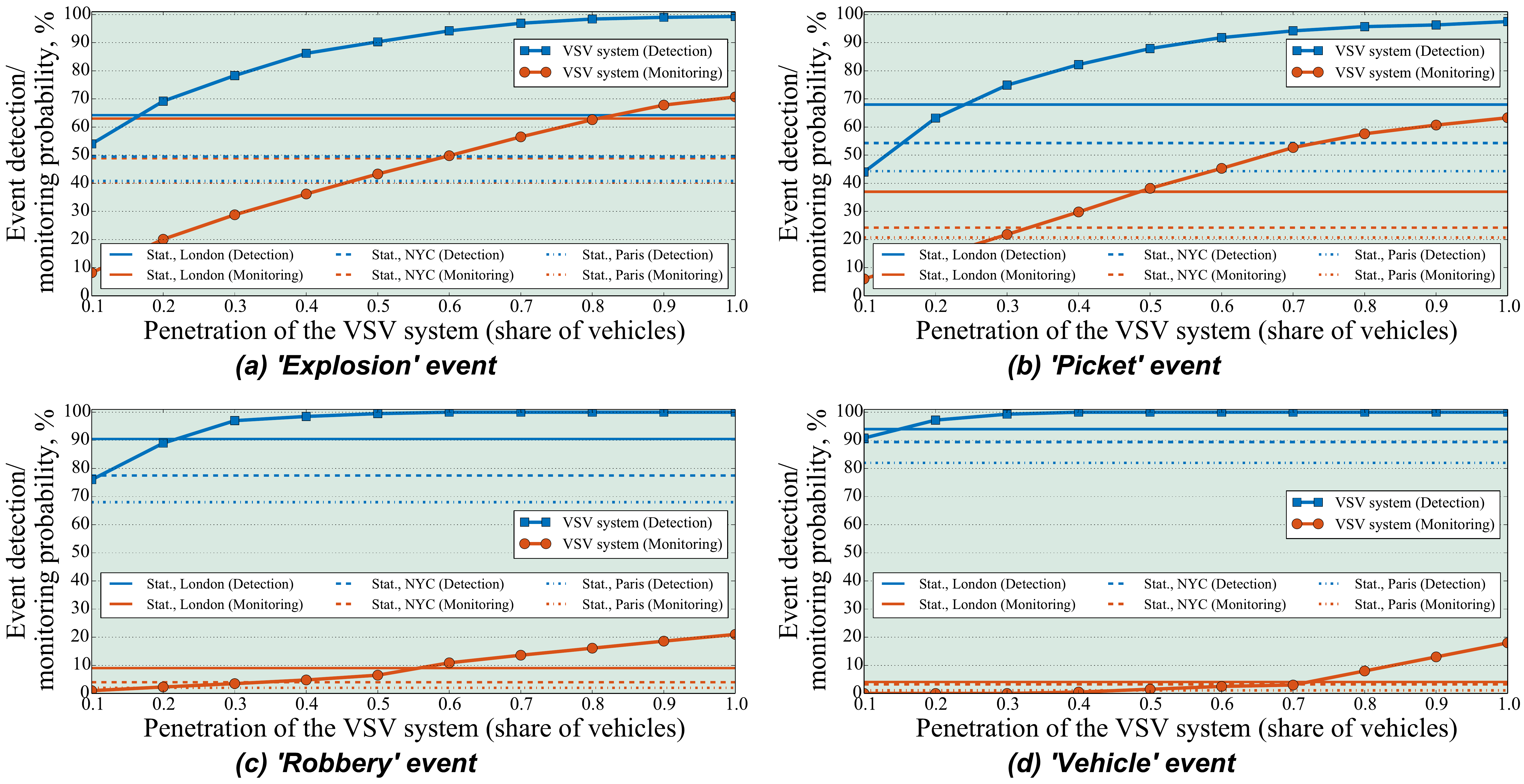}
   \vspace{-7mm}
  \caption{Detection and monitoring probabilities for different categories of events. VSV system vs. stationary video surveillance setup deployed with camera density as in London, New York City (NYC), and Paris.}
  \label{fig:share}
   \vspace{-5mm}
\end{figure*}

In order to disregard the effects of particular road traffic flows, the vehicles are assumed to move with constant speeds and follow the Manhattan grid mobility pattern: at every intersection, the car randomly selects its new direction, except for the direction that it just came from. Any of the available new directions is chosen with equal probability. Since the primary focus of the presented case study is on the video monitoring service and following the discussion in Section~\ref{sec:concept}, optimistic assumptions about the radio links have been made. Other major parameters are summarized in Table~\ref{tab:parameters}.

\subsubsection{Simulation setup}

To model the mobility of both vehicles and events over a realistic path graph displayed in Fig.~\ref{fig:case_study}(b), an in-house simulation framework was constructed. The modeling process begins by identifying the area of interest with all the intersections and streets inside it, which are then converted into vertexes and edges of the path graph. All of the vehicles, stationary cameras, and the target events are deployed uniformly along the streets.

To obtain a first-order performance estimation for the described scenario, our in-house system-level simulator has been employed. The tool utilizes the vehicle mobility model discussed in the previous subsection to determine both location and rotation of the vehicles at any given moment of time. These data are then used to determine, which particular objects/zones on the map are currently being observed by the camera of a particular vehicle. Once a video fragment is captured, it introduces its time stamp and is to be transmitted to the application server, where it is later combined with other video fragments by following the procedure elaborated in Section~\ref{sec:concept}.

The simulation tool is fully implemented in Python and operates in a time-driven manner. The modeling step is set to $0.05$\,s, such that the vehicles can move for not more than $1$\,m between the timestamps. A single simulation round models one hour of real system operation. For the sake of improved accuracy, all of the collected intermediate data were averaged over $1,000$ independent rounds, thus altogether producing $1,000$\,hours of system operation.

\vspace{-1mm}
\subsection{Results and their Implications}
\vspace{-1mm}

\subsubsection{Comparing stationary vs. mobile}

First, Fig.~\ref{fig:share} reports on the probabilities to (i) detect a certain category of events and (ii) fully monitor them by either the proposed VSV system or the stationary video surveillance setup. The results are grouped individually for every event category.

Starting with the `Explosion' case presented in Fig.~\ref{fig:share}(a), we note that both the detection and the monitoring probability of a static event by stationary cameras are very similar. On the contrary, the VSV system output is different, since it has both very high chances to capture the event as well as decent chances to miss a part of it in space or time domain due to the mobility of cameras. Consequently, the VSV system is preferred in terms of the event detection probability starting from already $0.2$ of vehicle penetration, whereas it outperforms the stationary London case for the monitoring probability only after $0.8$ of vehicle penetration.

The results for the `Picket' category, see Fig.~\ref{fig:share}(b), are mostly similar to the previous case, except for two aspects: (i) detection and monitoring probabilities for the stationary case are not anymore the same due to mobility of the `Picket' event and (ii) our VSV system outperforms its stationary counterpart in terms of the monitoring probability already after $0.5$ of penetration. A similar trend continues for both the `Robbery' and the `Vehicle' categories, as shown in Fig.~\ref{fig:share}(c) and Fig.~\ref{fig:share}(d), respectively. This confirms that the VSV system may also be attractive for detecting events that are mobile.

\emph{Summarizing, the VSV system can effectively complement already deployed stationary solutions in certain aspects, while the combination of both stationary and mobile systems makes the detection of the considered events extremely reliable.}

\subsubsection{Video streams will be highly fragmented}

\begin{figure}[!b]
  \centering
    \vspace{-5mm}
  \includegraphics[width=0.9\columnwidth]{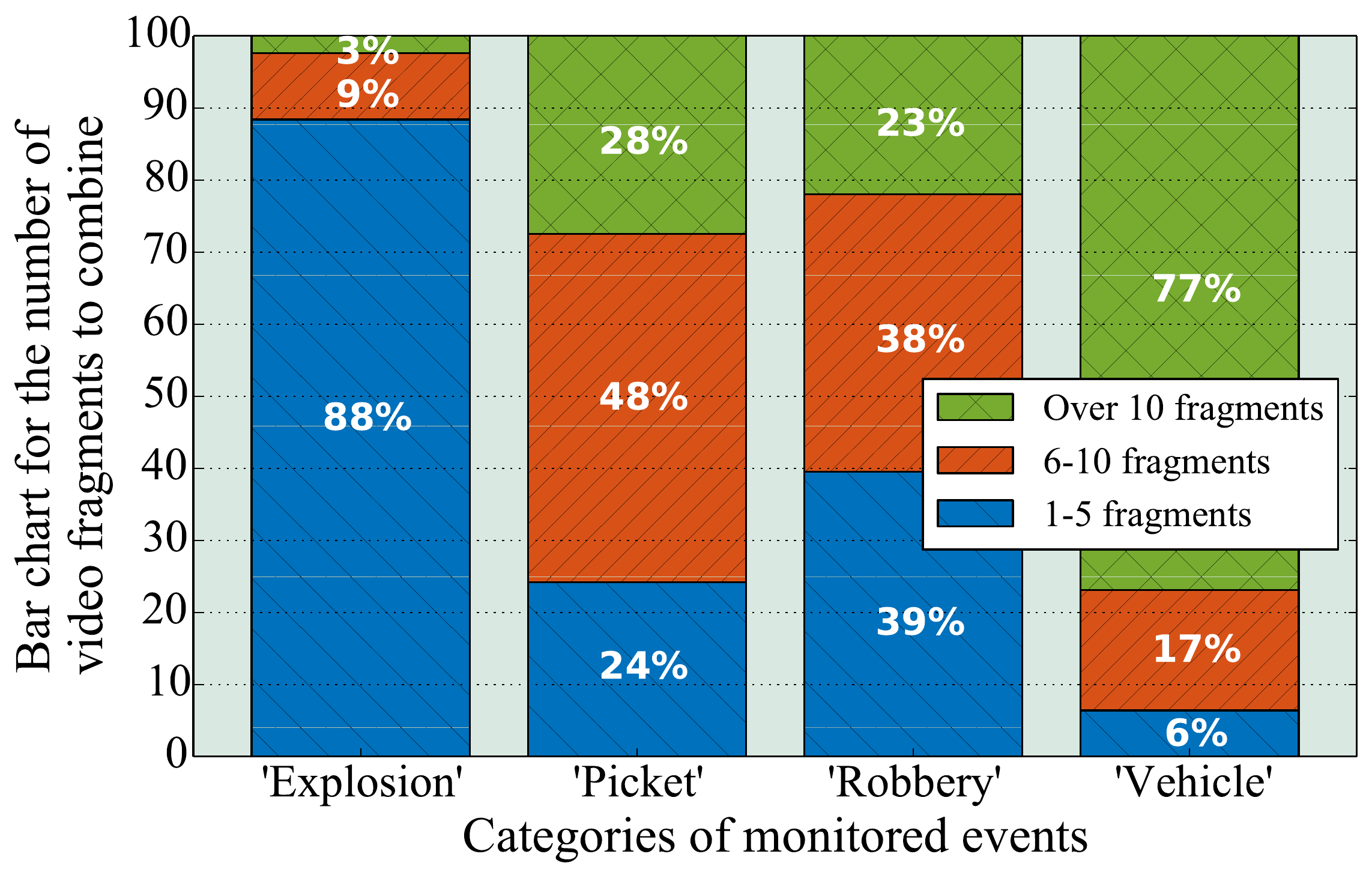}
  \vspace{-4mm}
  \caption{Level of fragmentation in collected video data assuming $30\%$ penetration of vehicles. Ordered based on event mobility speed.}
  \label{fig:usecases}
\end{figure}

Meanwhile, continuous event monitoring with the VSV system may be challenging. In Fig.~\ref{fig:usecases}, we observe to what extent monitoring of a certain event may be fragmented depending on its category. High numbers of the video fragments in the stream increase the complexity of combining them as well as that in recognizing the events. This also demands more resources to store data and perform search. As one may learn from the figure, the stream related to the two short-term cases (`Explosion' and `Robbery') is typically constructed by a moderate number of fragments. In contrast, the `Robbery' stream has over $20\%$ chances to exceed $10$ fragments, which is similar to values observed for a longer-term but less mobile `Picket' category. 

Comparing the results of Fig.~\ref{fig:usecases} with those offered in Fig.~\ref{fig:share}, we can also notice correlation between the difficulty to have a certain event fully monitored and the number of fragments in the resulting stream. Hence, even in the rare cases when a `Robbery' or a `Vehicle' event is fully captured by the VSV system, the stream consists of many short-term fragments, which may complicate the analysis. We discuss this and other important research challenges in the following section, as we introduce enablers for the VSV systems development.

\section{Research and Technology Challenges in VSV}
\label{sec:challenges}

The proposed VSV system poses novel research and engineering challenges that deserve special attention. These considerations and possible ways to resolve them are reported in this section. We introduce them category by category, thus forming an initial roadmap for the VSV systems development.

\vspace{-3mm}
\subsection{Connectivity}

While a number of alternative solutions exist for wireless communication with the vehicle, the choice of a particular option should be made carefully, since it has implications on the VSV service quality and the overall network performance.

\subsubsection{Transmitting alerts via cellular}

Most cities are already covered with the Long Term Evolution (LTE) networks, while cellular modules become a de-facto technology for the high-end vehicles of today as well as penetrate into the lower-price market segments. Going further, the use of NarrowBand IoT (NB-IoT) can be considered in certain cases to ensure scalability of the envisaged VSV system, offload the increasingly congested LTE networks, and provide reliable reporting of alerts in environments with limited LTE coverage.

\subsubsection{Scaling video transmission with mmWave}

At the same time, the capacity of LTE and NB-IoT solutions is absolutely insufficient to support massive video transmissions from a large number of connected vehicles in crowded urban scenarios. Hence, the utilization of capacity-friendly technologies, such as mmWave communication~\cite{heath_mmwave}, needs to be considered.

The use of mmWave small cells to provide wireless access for connected vehicles opens several research problems, primarily due to the high moving speeds of vehicles and short contact times between the vehicle and its serving small cell~\cite{jsac_SDN}. On the other hand, the mobility of a vehicle is much more predictable than that of a handheld user device. Therefore, predictive beam tracking and handover solutions may be employed to mitigate connection drops~\cite{heath_vehicular_sensing}. Simultaneous links to small cells realized via multi-connectivity techniques should also contribute to higher connection reliability levels~\cite{jsac_MC}.

\vspace{-3mm}
\subsection{Networking}

Further challenges arise at the upper layers of the protocol stack in the VSV system. Due to high speeds of the end nodes and large volumes of data to be collected, the envisioned VSV architecture should not rely exclusively on static (even multi-radio) infrastructures for wireless connectivity. Instead, the use of complementary solutions has to be considered.

\subsubsection{Harnessing V2V for VSV}

First, the utilization of direct vehicle-to-vehicle (V2V) links has the potential to offload certain amounts of data from the static radio access networks. In this regard, two major options may be addressed. In the scenarios with high relative velocity, such as a city, connected vehicles featuring mmWave V2V interfaces may form a type of ``train'', wherein the car currently connected to the mmWave small cell acts as a gateway to transmit the aggregate data. In these use cases, the gateway role may pass between the vehicles frequently, which in its turn raises challenges along the lines of addressing, mobility management, and handover.

By contrast, in the scenarios with low relative velocity, such as a highway, certain vehicles may be identified as gateways for longer periods of time, since the set of proximate cars changes relatively infrequently. Accordingly, the gateway vehicle may collect information received from the end nodes and, whenever there is an opportunity, inject such aggregate data into the access network (e.g., via a mmWave small cell on its way). The described approach results in some form of a hierarchy on top of the connected vehicles, thus ultimately forming a so-called \emph{vehicular cloud network}~\cite{mario_vehicular_cloud}.

\subsubsection{Keeping edge cloud busy}

The advantages brought by the edge cloud should also be studied carefully. In particular, different approaches may be utilized to process the received data. If restricted by its connectivity or maximum data rate, the intermediate node can: (i) delay the data transmission in case the network is currently overloaded; (ii) distribute the received data between several processing units for improved load balancing; and (iii) perform a certain level of pre-processing locally if equipped with the relevant computing capabilities. Each of the discussed options has several degrees of freedom, thus resulting in specific performance tradeoffs. Hence, the networking aspects of VSV become a key building block for the overall system implementation.

\vspace{-3mm}
\subsection{VSV Data Storage and Processing}

While the general principles behind VSV data processing were outlined in Section~\ref{sec:concept}, the choice of specific solutions to conduct certain actions have important implications on the system performance and thus deserve special attention.

\subsubsection{Structuring the mess}

First and foremost, there is a need for an effective solution to store the obtained heterogeneous information. Here, the tradeoffs enabled by the utilization of particular relational or non-relational databases, where different categories of data are stored and indexed separately, must be clearly identified. For instance, a relational database may be utilized to store the detected events, while a NoSQL type of storage could be adopted for the video content.

\subsubsection{Looking for a needle in a haystack}

Further, the stored and combined data have to be processed to produce valuable knowledge for the service consumer. Such massive amounts of multimedia big data cannot be handled by human beings and may comprise numerous minor details that people are unlikely to even notice. Therefore, the respective application of automated data mining methods is envisioned to bring decisive benefits across many industries. Here, the use of the most advanced solutions from computer vision, high-order correlation analysis, and deep learning has to be attempted~\cite{video_pattern}.

\vspace{-2mm}
\subsection{Security and Privacy Concerns}
\vspace{-1mm}

The emergence of VSV systems requires addressing a number of important security-related considerations as well as accentuates the critical privacy-centric ethical questions.

\subsubsection{Big data -- big attraction}
Sensitive big data attracts increased attention of attackers, while numerous potential risks arise in such a distributed ecosystem. Therefore, highly-secure chains of trust have to be constructed and maintained in real-time. The application of pre-configured approaches may mitigate the risks in early VSV deployments. Later on, presently available security techniques to increase the VSV system flexibility (dynamic chains of trust, white/black lists, ID-based cryptography, etc.) need to be revised.

\subsubsection{Assistance from hardware}
In its turn, the envisaged system for partial data pre-processing in the vehicle/cloud nodes is difficult to implement in a secure manner (such that the pre-processing logic remains trusted by other network nodes) without certain hardware-level solutions. In particular, the use of Trusted Execution Environment (TEE)~\cite{nokia_tee} can ensure the authenticity of software at all stages of the VSV system workflow. In addition, TEE can simplify certain security-oriented functions (e.g., encryption of large data volumes). Hence, seamless integration of hardware- and software-based techniques is another key challenge to be faced.

\subsubsection{Privacy-centric access control}
Our proposed approach should maintain the capability to differentiate between the types of data to be shared with various consumers. For instance, a policy may be introduced that restricts regular consumers to only access the video data, while the law enforcement units may also decrypt the identifiers of the source vehicles (e.g., to question the drivers as potential witnesses). One of the solutions here is to apply asymmetric encryption to protect the identifiers. In case the public key of the police is used to encrypt the identifiers, only appropriately authorized law enforcement units will be able to decrypt the IDs.

\subsubsection{Privacy-preserving data collection}
For the sake of preserving privacy, the source of a collected data fragment needs to be kept anonymous for most of the customers; hence, appropriate data anonymization techniques should be applied~\cite{extra_privacy}. Here, encryption of the vehicle identifiers alone is not sufficient, since one is still capable of tracing the video fragments associated with the same encrypted ID. One of the options to address this problem is to append a timestamp and a random number (nonce) to the identifier before encryption. Therefore, the encrypted ID will be different for various video fragments and only an authorized entity will be able to link together all of the video data captured by a certain vehicle.

\subsubsection{Privacy provisioning for drivers and pedestrians}
Privacy of pedestrians and other human actors with respect to their appearance on the video has to also be protected. The sensitive parts of the video feed (faces, plate numbers, etc.) may be recognized automatically and then blurred at the vehicle side before transmission. By combining the described approaches, a regular VSV service provider will only operate with anonymized data, while any access to full data (including the video without the blur and the IDs) is to be provided exclusively to a limited number of authorized stakeholders.


\balance
\section{Conclusions}
\vspace{-1mm}
\label{sec:conclusions}

Crowd sourced video surveillance over vehicles is envisioned to complement the existing niche solutions that cover a certain area of interest with stationary cameras. By design, the proposed system offers better scalability and greater flexibility versus such static setups, as well as provides higher predictability against video surveillance with handheld devices. At the same time, our VSV system opens the door to deep cross-integration of different clusters, including automotive, telecommunication, data mining, and government. It also transforms the data captured and collected in one cluster to infer valuable knowledge for the other, thus generating benefits for all the involved stakeholders. Meanwhile, a set of novel research challenges emerge, calling both for new mechanisms to manage and interpret the collected fragmented data as well as for advanced approaches to build the associated networking infrastructure for crowd sourced multimedia big data in challenging urban environments.

\section*{Acknowledgment}
This work was supported by the Academy of Finland (project PRISMA) and by the project TAKE-5: The 5th Evolution Take of Wireless Communication Networks, funded by Tekes. The work of S. Andreev was supported in part by a Postdoctoral Researcher grant from the Academy of Finland and in part by a Jorma Ollila grant from Nokia Foundation. The work of V. Petrov was supported in part by a scholarship from Nokia Foundation and in part by HPY Research Foundation funded by Elisa.

\vspace{-1mm}
\bibliographystyle{ieeetr}
\bibliography{vehicles_video}

\begin{thebibliography}{10}

\bibitem{ref_video}
T.~Y. Wu, S.~Guizani, W.~T. Lee, and P.~C. Huang, ``An enhanced structure of
  layered forward error correction and interleaving for scalable video coding
  in wireless video delivery,'' {\em IEEE Wireless Communications}, vol.~20,
  pp.~146--152, August 2013.

\bibitem{IHS_Markit}
I.~Markit, ``{Top Video Surveillance Trends for 2017},'' {\em White Paper},
  2017.

\bibitem{discomfort_book}
F.~Bjorklund and O.~Svenonius, {\em {Video Surveillance and Social Control in a
  Comparative Perspective}}.
\newblock Routledge, 2013.

\bibitem{heath_vehicular_sensing}
J.~Choi, V.~Va, N.~Gonzalez-Prelcic, R.~Daniels, C.~R. Bhat, and R.~W. Heath,
  ``Millimeter-wave vehicular communication to support massive automotive
  sensing,'' {\em IEEE Communications Magazine}, vol.~54, pp.~160--167,
  December 2016.

\bibitem{related_work2}
P.~Gomes, C.~Olaverri-Monreal, and M.~Ferreira, ``Making vehicles transparent
  through {V2V} video streaming,'' {\em IEEE Transactions on Intelligent
  Transportation Systems}, vol.~13, pp.~930--938, June 2012.

\bibitem{related_work3}
A.~Vinel, E.~Belyaev, K.~Egiazarian, and Y.~Koucheryavy, ``An overtaking
  assistance system based on joint beaconing and real-time video
  transmission,'' {\em IEEE Transactions on Vehicular Technology}, vol.~61,
  pp.~2319--2329, June 2012.

\bibitem{ref_camera}
T.~E. Dokor, ``Autonomous vehicles need in-cabin cameras to monitor drivers,''
  {\em IEEE Spectrum}, October 2016.

\bibitem{mario_vehicular_cloud}
E.~Lee, E.~K. Lee, M.~Gerla, and S.~Y. Oh, ``Vehicular cloud networking:
  {A}rchitecture and design principles,'' {\em IEEE Communications Magazine},
  vol.~52, pp.~148--155, February 2014.

\bibitem{mario_incentives}
S.~B. Lee, J.~S. Park, M.~Gerla, and S.~Lu, ``Secure incentives for commercial
  ad dissemination in vehicular networks,'' {\em IEEE Transactions on Vehicular
  Technology}, vol.~61, pp.~2715--2728, July 2012.

\bibitem{heath_mmwave}
J.~G. Andrews {\em et~al.}, ``Modeling and analyzing millimeter wave cellular
  systems,'' {\em IEEE Transactions on Communications}, vol.~65, pp.~403--430,
  January 2017.

\bibitem{jsac_SDN}
V.~Petrov {\em et~al.}, ``Achieving end-to-end reliability of mission-critical
  traffic in softwarized 5{G} networks,'' {\em IEEE Journal on Selected Areas
  in Communications}, vol.~36, pp.~485--501, March 2018.

\bibitem{jsac_MC}
V.~Petrov {\em et~al.}, ``Dynamic multi-connectivity performance in ultra-dense
  urban mm{W}ave deployments,'' {\em IEEE Journal on Selected Areas in
  Communications}, vol.~35, pp.~2038--2055, September 2017.

\bibitem{video_pattern}
A.~Barbu, Y.~She, L.~Ding, and G.~Gramajo, ``Feature selection with annealing
  for computer vision and {B}ig {D}ata learning,'' {\em IEEE Transactions on
  Pattern Analysis and Machine Intelligence}, vol.~39, pp.~272--286, February
  2017.

\bibitem{nokia_tee}
J.~E. Ekberg, K.~Kostiainen, and N.~Asokan, ``The untapped potential of
  {T}rusted {E}xecution {E}nvironments on mobile devices,'' {\em IEEE Security
  \& Privacy}, vol.~12, pp.~29--37, July 2014.

\bibitem{extra_privacy}
L.~A. Dunning and R.~Kresman, ``Privacy preserving data sharing with anonymous
  id assignment,'' {\em IEEE Transactions on Information Forensics and
  Security}, vol.~8, pp.~402--413, February 2013.

\end{thebibliography}

\end{document}